\documentclass[prl,amsmath,amssymb,reprint, superscriptaddress,nofootinbib,floatfix]
{revtex4-1}
\usepackage{amsmath,amssymb}
\usepackage{hyperref}
\usepackage{graphicx}
\usepackage{import}
\usepackage{tikz}

\newcommand{\diff}{\text{d}}

\begin{document}
\title{Periodic orbit theory of Bethe-integrable quantum systems: \\
an $N$-particle Berry-Tabor trace formula}
\newcommand{\RegensburgUniversity}{Institut f\"ur Theoretische Physik, 
Universit\"at Regensburg, D-93040 Regensburg, Germany}

\author{Juan Diego Urbina}
\affiliation{\RegensburgUniversity}
\author{Michael Kelly}
\affiliation{\RegensburgUniversity}
\author{Klaus Richter}
\affiliation{\RegensburgUniversity}
\begin{abstract}
One of the fundamental results of semiclassical theory is the existence of trace formulae showing how spectra of quantum mechanical systems emerge from massive interference among amplitudes related with time-periodic structures of the corresponding classical limit. If it displays the properties of Hamiltonian integrability, this connection is given by the celebrated Berry-Tabor trace formula, and the periodic structures it is built on are KAM tori supporting closed trajectories in phase space. Here we show how to extend this connection into the domain of quantum many-body systems displaying integrability in the sense of the Bethe ansatz, where a classical limit cannot be rigorously defined due to the presence of singular potentials. Formally following the original derivation of Berry and Tabor 
\cite{Berry1976,Berry1977}, but applied to the Bethe equations without underlying classical structure, we obtain a  many-particle trace formula for the density of states of $N$ interacting bosons on a ring, the Lieb-Liniger model. Our semiclassical expressions are in  excellent agreement with quantum mechanical results
 for $N=2,3$ and 4 particles. For $N=2$ we relate our results to the quantization of billiards with mixed boundary conditions. Our work paves the way towards the treatment of the important class of integrable many-body systems by means of semiclassical trace formulae pioneered by Michael Berry in the single-particle context.
\end{abstract}
%

%
\maketitle

\section{I. Introduction}

Among Michael Berry's many outstanding achievements, covering so many facettes of Theoretical Physics ranging from quantum optics to condensed matter, asymptotic analysis takes a particularly prominent role and deserves special mention \cite{Berry1972}. 
A key aspect of this way of thinking, where the limiting procedure involves a proper treatment of singularities, comprises emergent phenomena in the crossover regime where quantum mechanics can be formulated in terms of underlying classical structures. In this semiclassical regime, the sophisticated combination of $\hbar$ with classical functions, defined over invariant manifolds of the corresponding classical Hamiltonian system, allows for the construction of quantum  mechanical transition amplitudes and spectral properties \cite{Berry1981,Gutzwillerb}.

The use of semiclassical methods has opened the door to a precise formulation of quantum signatures of chaos~\cite{Haakeb} for quantum systems with a classical limit. In this semiclassical regime characteristic properties of quantum spectral and dynamical properties are intimately linked to chaos or integrability on the classical side. The history of the field, earlier also coined quantum chaology~\cite{Berry1989}, encompasses the development of these correspondences, starting from first phenomenological observations to the advanced level that has been achieved during the last 20 years in single-particle~\cite{Haakeb} and many-body semcilassical physics~\cite{Richter2022}. For the development of this semiclassical program, Michael Berry's contributions have been foundational.
In this respect, the discovery of the famous semiclassical trace formula for systems with integrable classical dynamics in 1976 \cite{Berry1976} and 1977 \cite{Berry1977} by M. Tabor and M. Berry is a prime example.
Before briefly reviewing this approach from a nowadays perspective, it is worth to mention their scientific context.
For systems with a fully chaotic classical limit, the corresponding foundational work by Gutzwiller~\cite{Gutzwillerb}
allows for expressing the density of states 
\begin{equation}
    \rho(E)=\sum_{n}(E-E_{n})
\end{equation}
of a quantum system described by the Hamiltonian $H(\hat{\bf p},\hat{\bf q})$
as the sum of two contributions,
\begin{equation}
\label{eq:rho2}
    \rho(E)=\bar{\rho}(E)+\tilde{\rho}(E) \, .
\end{equation}
For a system with $f$ degrees of freedom, the so-called smooth part 
\begin{equation}
    \bar{\rho}(E)=\frac{1}{(2\pi \hbar)^{f}}\int d{\bf p}d{\bf q}\delta(E-H({\bf p},{\bf q}))\, 
\end{equation}
is approximately given by the Thomas-Fermi approximation, or its variants and extensions, in terms of the phase space volume of the energy shell of the corresponding classical Hamiltonian $H({\bf p},{\bf q})$.
The second,  oscillatory contribution  $\tilde{\rho}(E)$ in Eq.~(\ref{eq:rho2}) is expressed as a sum over all classical periodic orbits at energy $E$, weighted by their stability and contributing with oscillatory amplitudes ${\rm e}^{iS_{\rm PO}(E)/\hbar}$. Here the ratio between their classical actions $S_{\rm PO}(E)$ and $\hbar$ gives rise to highly oscillatory phases in the semiclassical limit.
For chaotic quantum systems, access to individual energy levels (and eigenfunctions \cite{Berry1989}) requires a tour de force that began with the Gutzwiller trace formula and its convergence issues, all the way to resummed expressions over periodic orbit contributions
towards refined semiclassical representations of the quantal density of states. Thereby, for certain systems energy levels could be clearly identified and discerned, and once again Berry's work~\cite{Berry1990} represents a keystone in the elusive goal of quantizing chaos \cite{Berry1990}.

In the case of classically integrable systems the situation has been much clearer (see \cite{Tabor1989} for a precise definition of integrability in classical Hamiltonian systems). Here, following approaches dating back to the foundations of quantum theory, a set of phenomenological rules was successfully developed for obtaining quantized energies based on the existence of constants of motion, the hallmark of integrable motion. This theory was developed with great success, first for simplified models for atomic systems and then extended to general integrable systems.

In modern terminology, the refinement and extension of this program goes under the name of "torus-" or "Eisntein-Brillouin-Keller-" (EBK in short) quantization. It provides the set of implicit equations,
\begin{equation}
\label{eq:EBK}
    {\bf J}({\bf c})=\hbar\left({\bf m}+\frac{{\boldsymbol \alpha}}{4}\right) \, ,
\end{equation}
for the $f$ constants of motion ${\bf c}$ (one of them being the energy) in terms of the quantum numbers ${\bf m}$ and Maslov indexes ${\boldsymbol \alpha}$ counting the number of turning points along the corresponding loop. The actions
\begin{equation}
   J_{i}=\frac{1}{2\pi}\int_{\gamma_{i}({\bf c})}{\bf p} \cdot {\bf dq} 
\end{equation}
are defined by the $f$ irreducible loops on the phase space manifold, determined by fixing the set ${\bf c}$ and has the topology of an $f$-torus, with the $i$th loop passing through $\alpha_{i}$ classical turning points \cite{Ozorio1988}. 
Finally, the set of integers $m_{i}$ will provide the semiclassical quantum numbers labelling the spectrum of the system. For the historical development of these ideas, as well as the modern technical aspects see Refs.~\cite{Gutzwillerb, Tabor1989, Ozorio1988}.

This semiclassical approach to integrable systems was lacking a way to compute the density of states itself in the spirit of a trace formula, namely as a sum over certain types of classical invariant manifolds, without resorting to the one-by-one enumeration given by the torus quantization. This problem was addressed and beautifully solved by Berry and Tabor in Refs.~\cite{Berry1976}, \cite{Berry1977}.

In a nutshell, the Berry-Tabor trace formula arises from a careful Poisson summation of the discrete energies implicitly defined by the torus quantization conditions (\ref{eq:EBK}). Since Poisson summation introduces extra integrations, the key technical aspect comprises the evaluation of these integrals within stationary phase approximation, thereby adequately capturing the correct asymptotic behavior for classical actions much bigger than Planck's constant, a condition that is appropriate for energies not too close to the ground state. This program, as straightforward as it seems, faces mathematical complications and requires careful identification of the geometrical objects involved as achieved in Refs.~\cite{Berry1976,Berry1977}. In the spirit of the trace formulae, then the density of states is constructed out of the massive interference among progressively longer orbits labelled by their winding numbers around those tori that support periodic motion, the {\it topological sum}.

Given the trace formula for the quantization of classical integrable systems in the form of a summation over periodic paths, several fundamental questions concerning the spectral properties of integrable systems can be addressed. Among them, we mention the semiclassical interpretation of the Poisson-like behavior of short range spectral fluctuations. This  has large numerical and experimental support as the integrable systems' analogue of the Bohigas-Giannoni-Schmid conjecture for chaotic systems~\cite{Bohigas1984}. In Ref.~\cite{Berry1977b} Berry and Tabor provide a sketch of a proof of this finding, based on their trace formula. A second benchmark application of trace formula-based methods for integrable quantum systems is the semiclassical calculation of measures for (non-universal) long-range spectral fluctuations such as the spectral rigidity by Berry \cite{Berry1984}. 

By construction the Berry-Tabor trace formula relies on the existence of a well defined classical limit in terms of a Hamiltonian dynamics where the concept of classical Hamiltonian integrability can be unambiguously employed. Torus quantisation then relies on the semiclassical approximation to the energy levels in terms of quantised classical action variables whose existence is guaranteed by the definition of classical integrability. 
This requirement of classical integrability in this precise sense seems, however, too restrictive. This is evident for certain classes of quantum systems without a classical limit, which still possess both dynamical and spectral properties that one would naturally associate with integrable behaviour. Such systems are referred to as {\it quantum integrable}~\cite{Thaker1981}. Such quantum models
are also called "exactly solvable systems".
Here we consider the family of such quantum mechanical models that admit a solution by means of a Bethe ansatz \cite{Thaker1981,Korepin1993}, that we take as working definition for quantum integrability.
The subject of Bethe-ansatz-solvable systems is too large to attempt for a comprehensible review here; instead we  refer the reader to the classic reference \cite{Korepin1993}, reflecting the high level of mathematical sophistication. Methods of quantum integrability represent  now standard tools in subjects ranging from quantum gravity and string theory to random matrix and condensed matter theory.

For our present purpose, however, the key ingerdient of the mathematical formulation of quantum integrability is the fact that, by its very definition, it implies the existence of a set of generally transcendent equations, the Bethe equations. They  implicitly define the constants of motion of the system, ${\bf k}$ (usually refereed to as rapidities), in terms of a set of (integer or half-integer) quantum numbers ${\bf I}$ that label the eigenstates of the Hamiltonian. From the rapidities, physical constants of motion like energy and momentum can be constructed.

There is, however, a class of generic quantum integrable systems without correspondence to quantized classically integrable systems: quantum integrable systems that do not admit a classical limit. In particular, we refer to the class of one-dimensional (1D) many-body systems with two-particle contact interactions. These systems include 1D models widely used in condensed matter scenarios like Fermi-Hubbard models describing interacting electrons on a lattice among many others \cite{Korepin1993}.

Correspondingly, here we address the question whether it is possible to extend the semiclassical Berry-Tabor approach 
beyond its natural formulation in terms of quantized tori, namely to quantum systems without such a classical limit.
We will show  how to extend the construction leading to the Berry-Tabor trace formula to  many-body quantum integrable systems for the particular situation describing the quantum dynamics of bosonic degrees of freedom in 1D space and interacting through delta-like potentials, specifically the paradigmatic Lieb-Liniger (LL) model~\cite{Lieb1963,Lieb2008}.

\begin{figure}[b]
\centering
\includegraphics[ width=0.8\columnwidth]{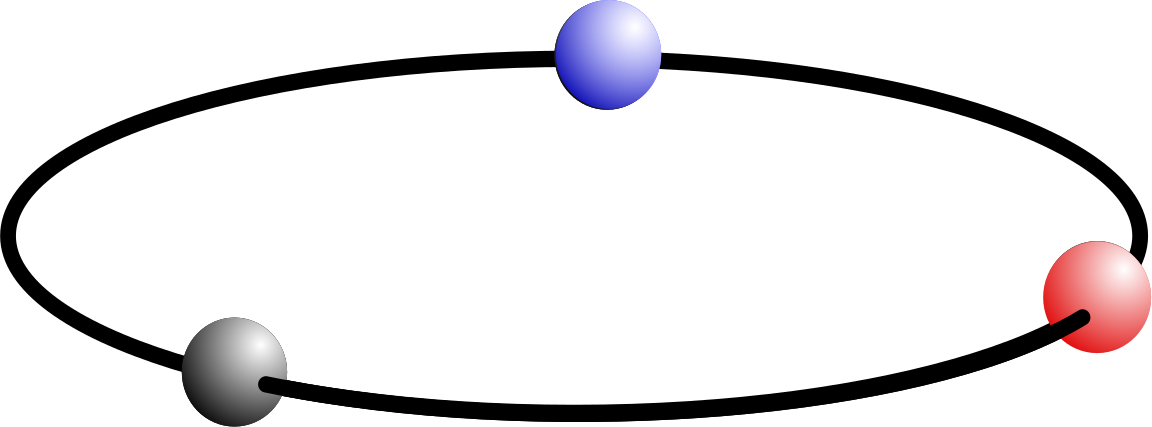}
\caption{The Lieb-Liniger model describing $N$ bosonic particles interacting through contact potentials and confined on a 1D ring}
\label{fig:fig1}
\end{figure}

\section{II. The Lieb-Liniger model}

 This famous model describes a 1D  integrable bosonic many-particle system with contact interaction on a ring of dimensionless length $L = 2
 \pi$.
 For $N$ particles with dimensionless coordinates $x_{i}$ 
 the Hamiltonian reads (in units of $\hbar= 2m = 1$) 
\begin{align}
	\mathcal{\hat{H}} = -\sum_{j=1}^N \partial_{x_j}^2 + 2g \sum_{1\leq k < j \leq N} \delta(x_j-x_k),
\end{align}
where $g$ is the interaction strength. In our work we focus on repulsive interactions, i.e., $g>0$. Finding the wavefunction and the eigenergies of the model is a well known task by using the Bethe ansatz \cite{Korepin1993}. Hence we will skip most of the details and just point out the key features for our purposes. 
Using the so-called coordinate form of the Bethe ansatz, and imposing the appropriate boundary conditions, one finds that the eigenergies can be expressed as 
\begin{align} 
\label{eq:EnergyLL}
	E(\vec{k}) = \sum_{j=1}^N k_j^2,
\end{align}
in terms of the parameters $k_j \in \mathbb{R}$ usually denoted as rapidities. They, in turn, are implicitly given by the Bethe equations
\begin{align} 
\label{eq:BetheEquations}
	2\pi I_j = Lk_j +2\sum_{i=1}^N \arctan\left(\frac{k_j-k_i}{g}\right).
\end{align} 
Here, the (half)integer numbers $I_j = m_j+\delta /2$, with $m_j \in \mathbb{Z}$ and $\delta = (N+1) \mod 2$ will provide the set of quantum numbers labelling the many-body eigenstates and referred to as Bethe quantum numbers.

We mention some key properties of the Bethe equations that will be important later, details can be found in the standard literature~\cite{Korepin1993}. First, by fixing a set of $N$ Bethe quantum numbers $\vec{I}$ to certain values, the rapidities $\vec{k}=\vec{k}({\vec I})$ are uniquely determined. Second, from Eq.~(\ref{eq:BetheEquations}) one can see that if two quantum numbers are equal the same has to be true for the rapidities, i.e. 
\begin{align} 
\label{eq:EqualActions}
 I_i = I_j \Rightarrow k_i = k_j.
\end{align}
Third, although the LL model refers to bosons, the excitations described by the eigenstates of the system are {\it fermionic}. Specifically, this means that, in order to get a non-vanishing wave function, all the Bethe quantum numbers $I_j$ have to be distinct from each other.
As a result of these three properties, and the fact that the total wavefunction must be symmetric under permutations of the particle's positions, we can restrict our analysis to the region $I_1<I_2<\ldots <I_N$, namely the fundamental domain. 
Finally, for the repulsive LL model, inspection of the Bethe equations (\ref{eq:BetheEquations}) indicates that there are two limiting cases of special character. On the one hand the so-called fermionized limit $g\rightarrow \infty$, also known as the Tonks-Girardeau gas. On the other hand, for $g=0$ the system simply reduces to free bosons as expected. 


\section{III: Many-body Berry-Tabor trace formula for the Lieb-Liniger model}

\subsection{The semiclassical expression for the spectral density}

In the following we derive a trace formula for the Lieb-Liniger model based on the Bethe equations. The whole approach is in close analogy to the work in Refs.~\cite{Berry1976,Berry1977}, where an EBK-quantization is used as starting point to construct a trace formula expressing the density of states of integrable models by a sum over classical periodic orbits.
The fact that the LL model, owing to its contact interaction, does not have a classical limit implies some interesting new features.

The density of states of the LL model can be expressed by a sum over the Bethe quantum numbers as
\begin{align}
\label{eq:rho}
	\rho(E) = \sum_{I_1<I_2<\ldots < I_N} \delta(E-E(\vec{k}(\vec{I})))
\end{align}
where the fermionic nature of the excitations implies an ordered sum over Bethe quantum numbers. This is a characteristic feature of this type of models that has no counterpart in the usual EBK quantization method.

If, following Berry and Tabor, we attempt to apply Poisson summation to Eq.~(\ref{eq:rho}), we must first overcome problems arising from the ordering of quantum numbers. This apparently simple task turns out to be quite a formidable one, without an explicit general solution. Nevertheless, a  formal expression can be obtained by using\footnote{As long as the function summed over is symmetric under the exchange of summation indices}
\begin{align}
\label{eq:order}
	\sum_{I_1<I_2<\ldots < I_N} \rightarrow \sum_{\vec{a} \in \Pi(N)} \left(C_{\vec{a}} \sum_{I_1,\ldots ,I_N} \sigma_{\vec{a}} \right)
\end{align}
with coefficients 
\begin{align}
\label{Hummel}
	C_{\vec{a}} =  \prod_{m=1}^N \frac{(-1)^{d(\vec{a})-N}}{m^{s_m} s_m!}
\end{align}
where $d(\vec{a})$ denotes the number of elements in the considered partition $\Pi(N)$ of the particle number $N$, and $s_m$ counts how often $m$ occurs in $\Pi$.  The sum over $\vec{a} \in 
\Pi(N)$ runs over all integer partitions ${\vec{a}}$ of $N$. The 
symbol $\sigma_{\vec{a}}$ is the product of Kronecker Deltas, 
\begin{align}
	\sigma_{\vec{a}} =& \delta(I_1,\ldots,I_{a}) \delta(I_{a_1+1,\ldots,I_{a_1+a_2}}) \ldots \nonumber\\
	&\times\delta({I_{\sum_i^{d-1}a_i+1},\ldots,I_{\sum_i^d a_i}}),
\end{align}
forcing some of 
the Bethe quantum numbers $I_i$ to be equal. Due to the identification of some of the quantum numbers 
$I_i$,  each of the correction terms on the RHS of Eq.~(\ref{eq:order}) has an effective dimension $d(\vec{a})$, where the actual values of $a_i$ indicate the number of coinciding  quantum 
numbers. 

While we checked the formula (\ref{Hummel}) by a brute-force evaluation of the multiple sums up to $N=10$, and a diagramatic type of proof can be found in \cite{Kato2001}, it is instructive to see its explicit form for $N=2,3$ where it reads 
\begin{align}
    \sum_{I_1 < I_2} \rightarrow \frac{1}{2!}
    \left( 
        \sum_{I_1, I_2} - \sum_{I_1 = I_2}
    \right),
\end{align}
and 
\begin{align}
    \sum_{I_1 < I_2 < I_3} \rightarrow \frac{1}{3!}
    \left( 
        \sum_{I_1, I_2, I_3} - 3\sum_{\substack{I_1, \\ I_2 = I_3}} + 2\sum_{I_1 = I_2 = I_3}
    \right).
\end{align}

The structure of the summation formula, Eqns.~(\ref{eq:order},\ref{Hummel}) can be understood as follows. By unrestricting the sum, we overcount terms for which some (or all) of the indices are the same, which then have to be subtracted. By doing so we  oversubtract terms which then have to be added again and so on. For the case of $N=2$ and $N=3$ this procedure is very straight forward. It gets more complicated as soon as there are more terms present with the same dimension $d(\vec{a})$ but different combinations of indices beeing equal.

Considering now Eq.~\eqref{eq:EqualActions}, these combinatorics imply for 
the energy, Eq.~\eqref{eq:EnergyLL},
\begin{align}
	E(\vec{k}) = \sum_{j=1}^{d(\vec{a})} a_j k_j^2
\end{align}
and for the rapidities in \eqref{eq:BetheEquations}
\begin{align}
\label{eq:BAa}
	2\pi I_j = L k_j + 2\sum_{i=1}^{d(\vec{a})} a_i \arctan\left(\frac{k_j-k_i}{g}\right).
\end{align}
Since the number $a_i$ counts the degree of degeneracy of the Bethe quantum numbers $I_i$, we call them in the following multiplicity factors. 

After this analysis, the density of states admits a corresponding  decomposition as
\begin{align}
 \rho(E) = \sum_{\vec{a} \in \Pi(N)} C_{\vec{a}}\rho_{\vec{a}}(E)
\end{align}
with the important consequence that now we can apply the Poisson summation to each of the terms, as they are given in terms of unordered sums. Denoting by $\vec{M}$ the new summation indices conjugated to the set (of now mutually different) ${\vec I}$ , we get
\begin{align} 
\label{eq:MthFluctuation}
	\rho_{\vec{a}}(E)=& \ \sum_{M_1, \ldots , M_{d(\vec{a})}} e^{i \pi \delta |M|}  \nonumber \\ 
	& \times \int \diff^d I \  \delta\left( E-E(\vec{k}(\vec{I}))\right)e^{2\pi i \vec{M} \cdot \vec{I}}\, ,
\end{align}
where $|M|=\sum_{i=1}^{d}M_{i}$
Here, the implicit relation  between the Bethe actions $\vec{I}$ and the rapidities $\vec{k}$ depends on the particular set $\vec{a}$ through the modified Bethe equations (\ref{eq:BAa}).
Terms with $\vec{M} = \vec{0}$ are fundamentally different from terms with $\vec{M} \neq \vec{0}$, as the exponential function vanishes and one is left with an integral over the energy contour only. As shown by Berry and Tabor for systems with classical limit this integral precisely gives the smooth Weyl contribution $\bar{\rho}$ to the density of states
(see Eq.~\ref{eq:rho2})) for systems with a classical limit. 
For LL-type systems with $N$ distinguishable and non-distinguishable  particles the mean density of states has been computed in Refs.~\cite{Hummel2014,Hummel2019}. 

Notably, for a quantum system without a genuine classical limit underlying the quantization conditions, as considered here, such term still provides the smooth background. 
Explicitely, by changing the integration variables $\vec{I}$ to the rapidities $\vec{k}$ we obtain for the smooth part
\begin{equation}
\label{eq:Weyl}
    \bar{\rho}(E)=\sum_{\vec{a} \in \Pi(N)} C_{\vec{a}}\int \diff^{d(\vec{a})} k \  \left|\frac{\partial \vec{I}}{\partial \vec{k}}\right|\delta\left( E-E(\vec{k})\right) \, ,
\end{equation}
in terms of the Jacobian $\partial \vec{I}/\partial \vec{k}$, related to the normalization of the Bethe states, and widely used in the formal approach to quantum integrable systems \cite{Korepin1993}. Equation~(\ref{eq:Weyl}) shows the power of the Berry-Tabor resummation, as all aspects of the model in the thermodynamic limit can be obtained from it. The non-trivial equivalence of $\bar{\rho}$ with the thermodynamic version of the Bethe ansatz was diagramatically shown in \cite{Kato2001}.

Beyond the smooth background provided by the Weyl term, the fine scale, eventually responsible for the $\delta$-peak structure of the density of states, is encoded in the fluctuating part with $\vec{M} \neq \vec{0}$. This allows us to decompose the entire density of states into $\bar{\rho}(E)$ and an oscillatory part, $\tilde{\rho}(E)$, i.e. 
\begin{equation}
	\rho(E) =  \bar{\rho}(E)+\tilde{\rho}(E)
	\\
	= \sum_{\vec{a} \in \Pi(N)} \bar{\rho}_{\vec{a}}(E)+\tilde{\rho}_{\vec{a}}(E) 
	.
\end{equation}
To solve the integral comprising the oscillatory contributions
we first switch again from $\vec{I}$ to $\vec{k}$ space, leading to
\begin{align}
\label{eq:Osc}
\tilde{\rho}_{\vec{a}}(E)=& \  \sideset{}{'}\sum_{M_1, \ldots , M_{d(\vec{a})}}  e^{i \pi \delta |M|}   \\ 
	& \times\int \diff^{d(\vec{a})} k \left| \frac{\partial \vec{I}}{\partial \vec{k}} \right| \delta\left( E-\sum_{j=1}^{d(\vec{a})} a_j k_j^2\right)e^{2\pi i \vec{M} \cdot \vec{I}(\vec{k})}, \nonumber
\end{align}
where the primed sum 
denotes that the term with $\vec{M} = \vec{0}$ is excluded. It is convenient to perform a fourier transformation and go from energy to time domain to get rid of $\delta$-distributions, i.e.
\begin{align}
\label{eq:full}
    \tilde{\rho}_{\vec{a}}(t) =& \frac{1}{2\pi}\int_0^\infty \diff E \ \tilde{\rho}_{\vec{a}} (E)\  e^{-iEt} \\
    =& \  \sideset{}{'}\sum_{M_1, \ldots , M_{d(\vec{a})}}  \frac{e^{i \pi \delta |M|}}{2\pi} \int \diff^{d(\vec{a})} k \left| \frac{\partial \vec{I}}{\partial \vec{k}} \right| e^{2\pi i \vec{M} \cdot \vec{I}(\vec{k}) - iE(\vec{k})t}. \nonumber
\end{align} 
Then we solve the remaining integral with stationary phase approximation. 
The corresponding stationary phase analysis results in the following set of equations:
\begin{align} \label{SPAfull}
	2\pi \vec{M} \cdot \frac{\partial \vec{I}}{\partial k_i} = t\ \frac{\partial E}{\partial \vec{I}} \cdot \frac{\partial \vec{I}}{\partial k_i} \, .
\end{align}
It contains the familiar 'periodic orbit' relation $\vec{M} \parallel \frac{\partial E}{\partial \vec{I}}$, first obtained in \cite{Berry1976}, and expresses the important fact that the trace formula is actually a sum over winding numbers characterizing topologically families of periodic orbits. Due to the similarity of our approach to that of Berry and Tabor the emergence of such condition is not  surprising. However, a precise interpretation of these 'periodic orbits' appears difficult since there is not an underlying classical phase space. 
 They can be viewed as being composed of segments of free many-particle propagation in between quantum scattering events of two colliding particles with contact interaction.
 Additionally, in contrast to integrable systems with classical limit the 'frequencies' $\partial E / \partial \vec{I}$ are not positive definite functions in the many-body case. 


\begin{figure}[ttt]
\centering
\includegraphics[ width=0.9\columnwidth]{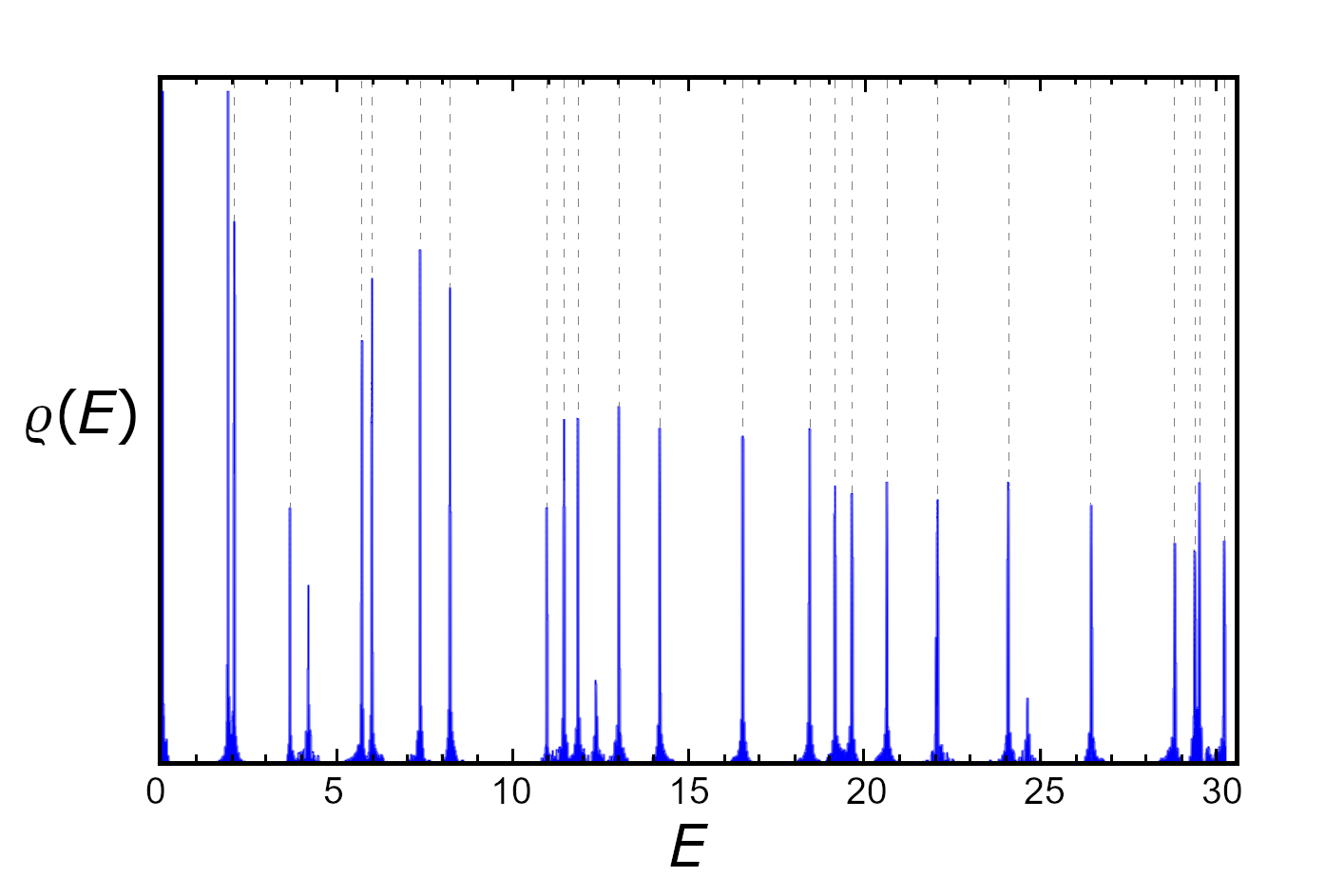}
\includegraphics[ width=0.9\columnwidth]{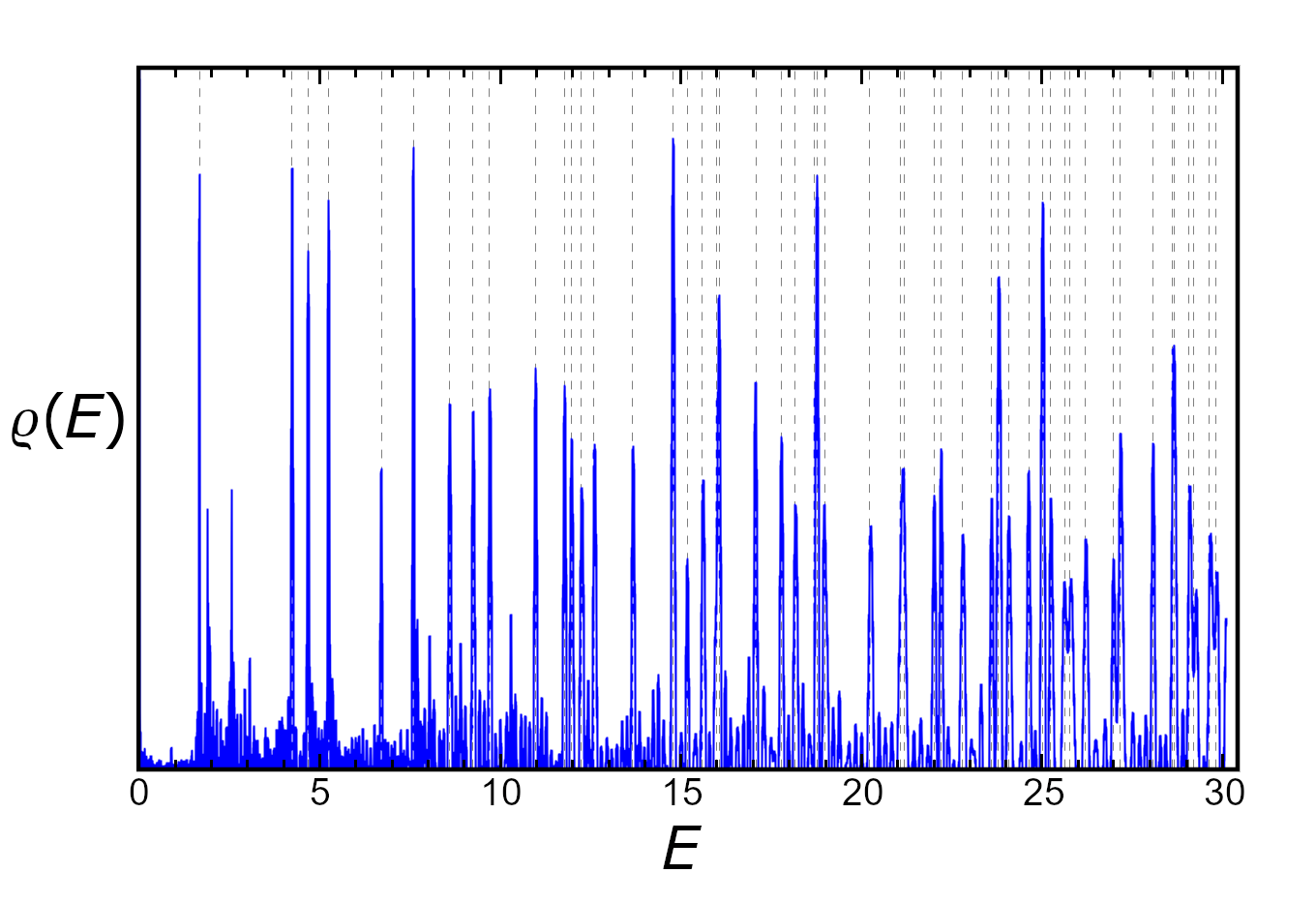}
\includegraphics[ width=0.9\columnwidth]{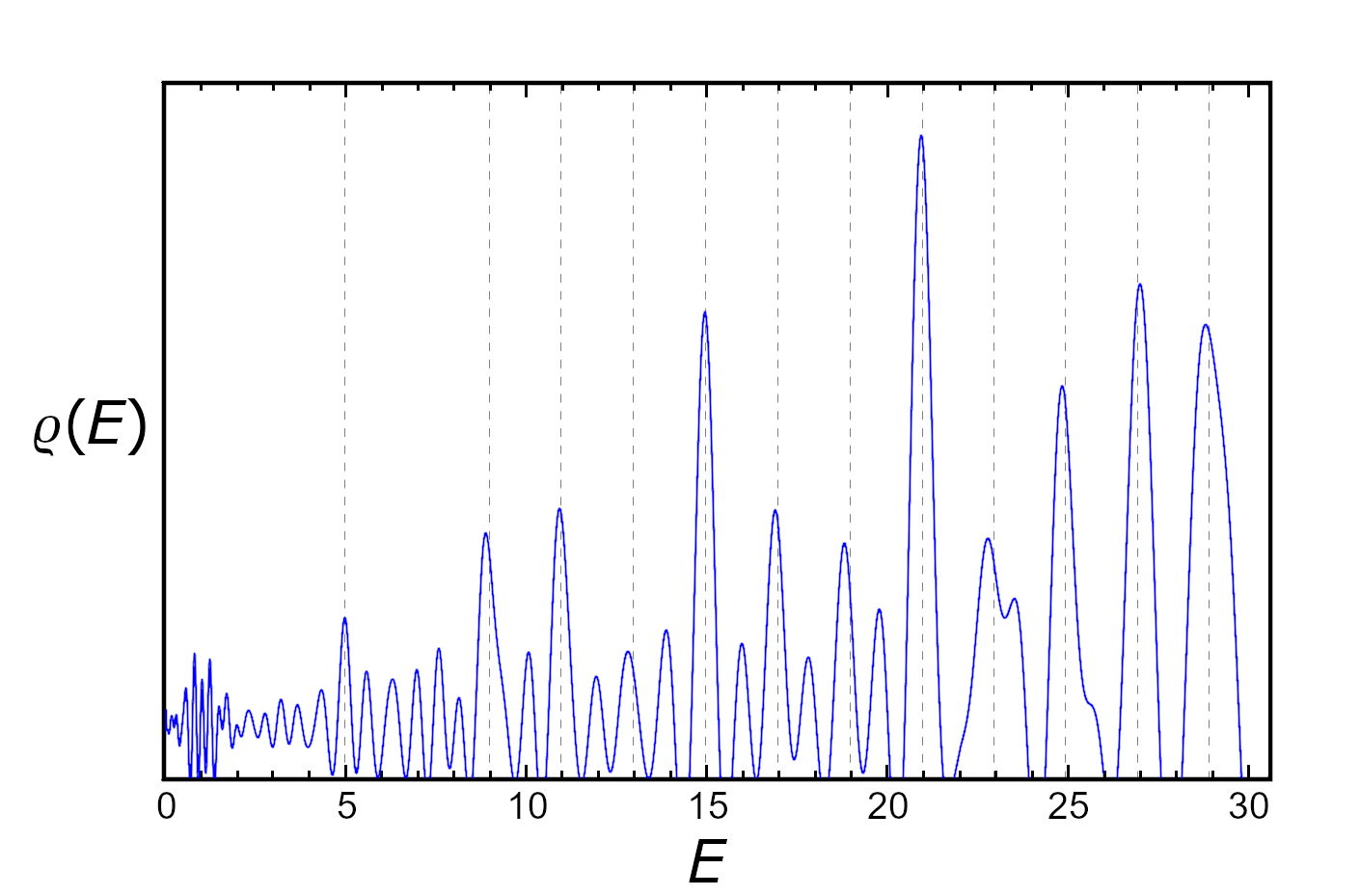}
\caption{Density of states $\rho(E)$ as function of the many-body excitation energy $E$ for a Lieb-Liniger system of $N=2,3$ and $N=4$ bosons (from top to bottom panel) in the strongly interacting regime $g=10$ and $g=100$ (bottom panel). Upon increasing the number of terms in the sum over winding numbers up to $M_{\textrm{max}}=10$ we see the emergence of discrete peaks located with high precision at the quantum mechanical many-body energy levels marked by vertical lines. (Energies are measured in units of $\hbar^{2}/2mL^{2}$.) Increasing $N$ for fixed $M_{\textrm{max}}$ diminishes the quality of the semiclassical approximation, but the energy levels are still clearly visible.
}
\label{fig:fig2}
\end{figure}

It would be now desirable to actually understand, based on the exact stationary conditions, the type of classical mechanics one can associate to such quantum integrable model, but since the Bethe equations are coupled transcendental equations one cannot solve Eq.~(\ref{SPAfull}) analytically. To proceed, we further assume the interaction part proportional to  $\arctan(k/g)$ to be small and exclude it from our stationary phase analysis, leaving a more precise study for future investigation. We want to stress however that, even at the level of this crude approximation, the effect of the quantum scattering due to the contact potential (and without classical limit) goes well beyond simply adding extra phases due to the interaction. In fact, quantum scattering strongly affects the {\it prefactors} themselves in a highly non-trivial manner due to the presence of the Jacobians $\partial \vec{I} / \partial \vec{k}$ in Eq~(\ref{eq:full}). The lack of an strict classical limit is a consequence of approaching the semiclassical regime within a first-quantization picture based on degrees of freedom being {\it particles}. In the thermodynamic limit $N \to \infty$, however, a different type of classical limit, given by the non-linear Schrödinger equation, emerges in the sense of {\it fields} \cite{Korepin1993}. Since contact potentials, specially in one dimension, are perfectly tractable  within classical (in general non-linear) wave equations, in this second-quantization approach the semiclassical program, including now a proper classical limit, can be formulated.

Coming back to the limit $g\rightarrow \infty$ of Eq.~(\ref{SPAfull}), in this so-called Tonks-Girardeau regime \cite{Korepin1993} the Berry-Tabor approach defines as classical limit the system of free fermions and uses this information to describe the model at arbitrary coupling $g$.  After going back from time to energy domain we obtain the following final expression for the oscillatory contribution to the many-body density of states in the Lieb-Liniger model:
\begin{align}
	\tilde{\rho}_{\vec{a}}(E) = \sideset{}{'}\sum_{M_1, \ldots , M_d}A_{\vec{M}}(E,\vec{a}) \times \cos(R_{\vec{M}}(E,\vec{a}))
 \label{eq:final}
\end{align}
with semiclassical amplitudes 
\begin{align}
	A_{\vec{M}}(E,\vec{a}) = \left(\frac{\pi^{d(\vec{a})-1}}{\prod_{i}^{d(\vec{a})} a_i}\right)^{\frac{1}{2}}\left(\frac{E^{d(\vec{a})-3}}{S_{\vec{M}}^{d(\vec{a})-1}}\right)^{\frac{1}{4}} \left| \frac{\partial \vec{I}}{\partial \vec{k}} \right|_{\vec{k}=\vec{k}_{\vec{M}}}
\end{align}
(note the presence of the Jacobians $\partial \vec{I} / \partial \vec{k}$) and actions
\begin{equation}
	R_{\vec{M}}(E,\vec{a}) \!=\!  \ 2\sqrt{S_{\vec{M}}E} \!-\! \frac{\pi}{4}(d(\vec{a})-1)
	\!+\! \vec{M}\cdot \vec{\Phi}_{\vec{M}}\!+\! \pi \delta |M|\, .
\end{equation}
%
%
Here
\begin{equation}
	S_{\vec{M}} = \sum_{i=1}^{d(\vec{a})} \frac{(LM_i)^2}{4a_i}  
	\quad ; \quad 
	(\vec{k}_{\vec{M}})_i = 
	\frac{LM_i}{2a_i} \sqrt{\frac{E}{S_{\vec{M}}}} \nonumber
\end{equation}
and
\begin{equation}
\label{eq:inter}
(\vec{\Phi}_{\vec{M}})_i = 2\sum_{j=1}^d a_j \arctan\left( \frac{ (\vec{k}_{\vec{M}})_i- (\vec{k}_{\vec{M}})_j}{g}\right) \, ,
\end{equation}
are easily recognized as the scaled actions and wavenumbers of free particles, and scattering phases between them. 

The set of equations (\ref{eq:final}-\ref{eq:inter}) is our main result. It allows for expressing the many-body density of states for the LL model as a sum where, starting with the smooth background term given by the generalized Weyl term $\bar{\rho}$ in Eq.~(\ref{eq:Weyl}), more and more structure is added in the form of Fourier-like oscillatory terms $\tilde{\rho}$ in Eq.~(\ref{eq:Osc}) associated with generalized periodic orbits.


\subsection{Comparison with quantum mechanical results}

To check the accuracy of our semiclassical prediction we compare results from Eq.~(\ref{eq:final}) with corresponding results from quantum mechanical solutions of the LL model. 
In Fig.~\ref{fig:fig2} we show how the peaks in the density of states obtained from the semiclassical trace formula (blue curves) approach the positions (dashed vertical lines) of the exact quantum few-body energy levels for a LL system with $N=2,3,4$ particles (top to bottom panel). As expected, for a given maximal winding number $M_{\textrm{max}}=10$ the quality of the semiclassical method decreases when increasing $N$, but the location of the few-body energy levels is clearly recovered. This emergence, however, does not occur in a level-by-level fashion, but rather by approaching  the $\delta$-peak structure of the exact density of states by means of massive interference.

\subsection{Resurgence in semiclassical few-particle spectra}

\begin{figure}[ttt]
\centering
\includegraphics[ width=0.9\columnwidth]{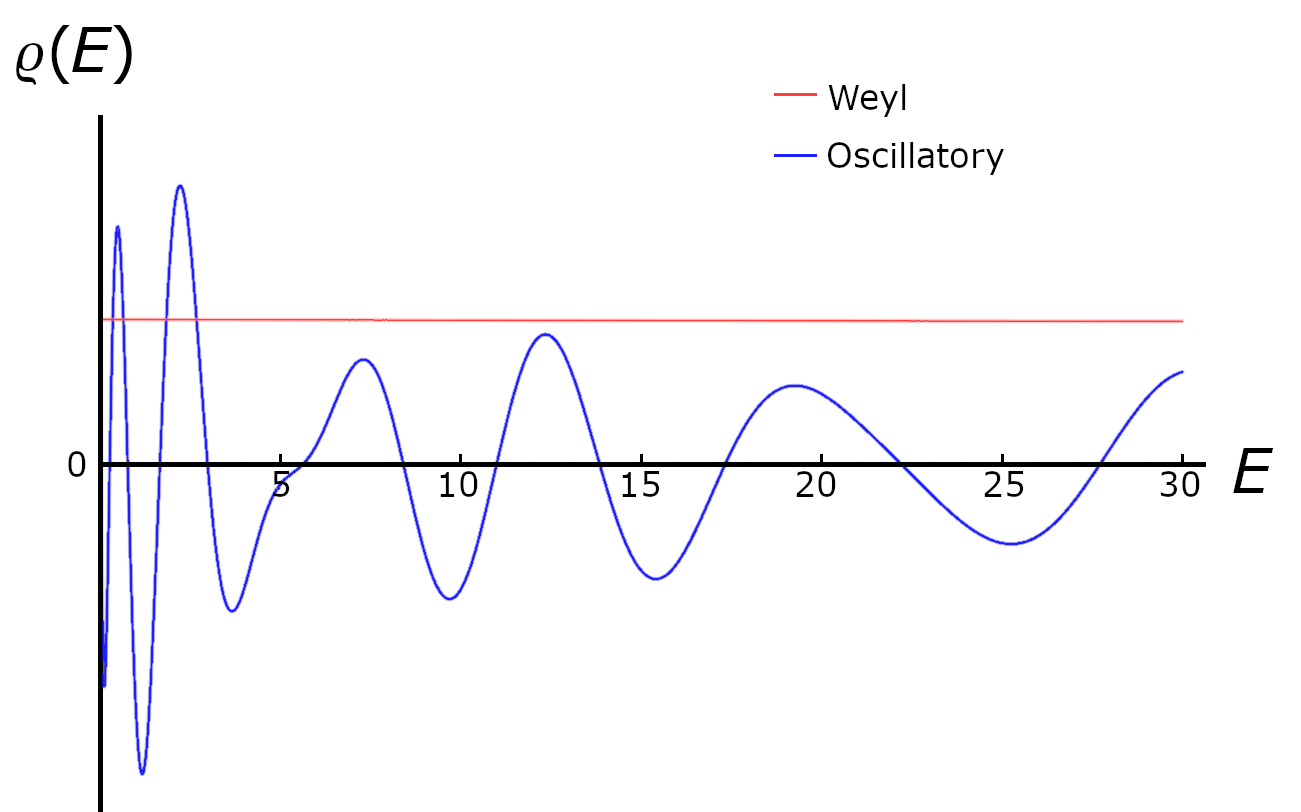}
\includegraphics[ width=0.9\columnwidth]{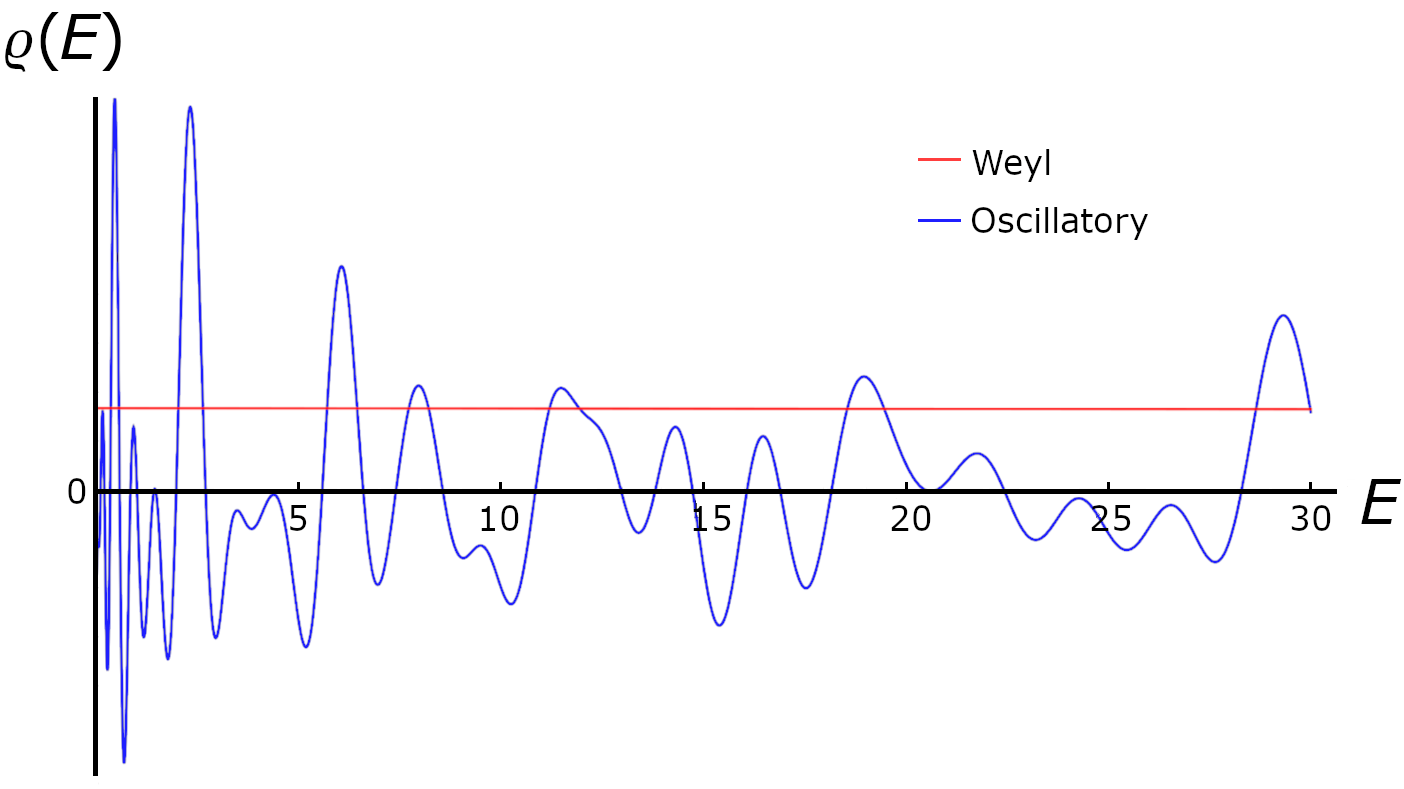}
\includegraphics[ width=0.9\columnwidth]{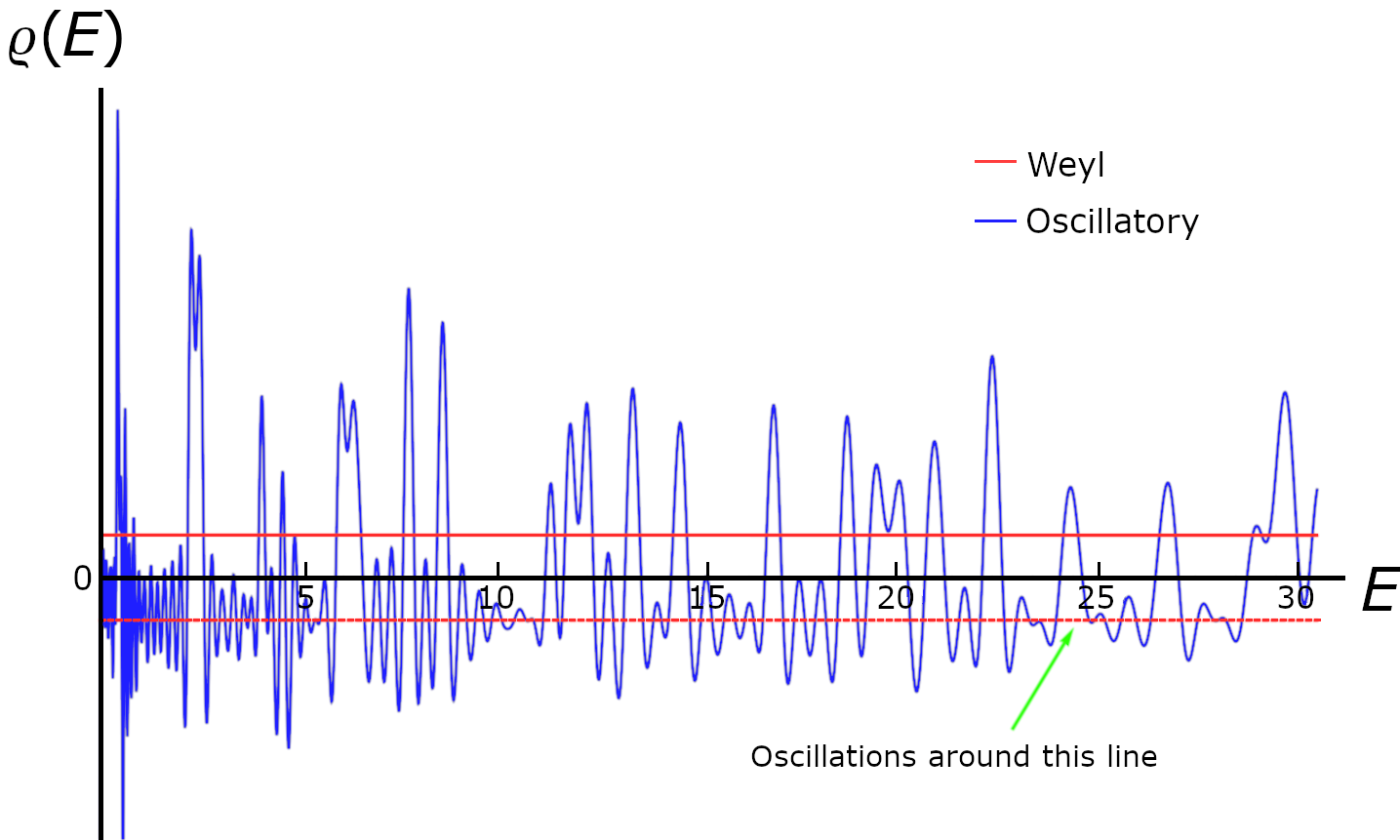}
\caption{Quantum mechanical resurgence, the cancellation of the smooth Weyl background  $\bar{\rho}(E)$ (red horizontal line) against an emergent negative background encoded in the oscillatory contribution, is clearly visible when increasing the number of terms in the periodic orbit expansion (here for maximal winding numbers $M_{\textrm{max}}=3,10,20$ from top to bottom). The resurgence phenomenon is demonstrated for the two-particle Lieb-Liniger model in the strongly interacting regime $g=10$. Energies are measure in units of $\hbar^{2}/2mL^{2}$.}
\label{fig:fig3}
\end{figure}

The high control we have for $N=2$ allows for studying a remarkable phenomenon that is clearly visible  in Fig.~\ref{fig:fig3} given the clear distinction between smooth and oscillatory contributions to the trace formula, Eq.~(\ref{eq:rho2}). With increasing $M_{\textrm{max}}=3,10,20$ (from top to bottom) and, correspondingly, an increasing  number of terms in the trace formula it is apparent how the smooth term is progressively cancelled by an emergent background stemming from the oscillatory contribution. This phenomenon was firstly reported by Berry \cite{Berry1989} for 1D systems of non-interacting particles, and is referred to as {\it resurgence}. 
From a purely mathematical perspective, resurgence is a trivial consequence of the structure of the trace formula with its characteristic form of a smooth term that is successively corrected by oscillatory periodic-orbit contributions. Since the asymptotic form of the semiclassical density of states must merge into a train of $\delta$-peaks, the oscillatory contribution must cancel the Weyl term, while at the same time forming the peaks. From a physical point of view one can also argue that, being responsible for the emergence of well defined peaks, the oscillatory contribution carries also the information about the low-resolution features of the spectrum, as its mean spectral density given by the Weyl term. These interpretations get, however, a non-trivial meaning when  considering that they are only strictly valid when the decomposition of the density of states as the sum of smooth and oscillatory terms is {\it exact}. Since our trace formula involves two key approximations, the cancellation of the smooth term by the resurgent sum over periodic orbits constitutes a stringent test of the approach presented here.


\subsection{Connection to billiard systems}

The case of a LL ystemw with $N=2$ that we just addressed can also be seen as a particular realization of an analogy between 1D interacting many-body systems of the LL type and the paradigmatic example of systems with a well controlled classical integrable limit, namely the particle in a box with $d=N$ dimensions. For this class of systems, the classical dynamics is defined by segments of free propagation inside the billiards's domain, connected by elastic reflections at the boundaries. These are textbook examples of Hamiltonian integrability as the squared linear momentum in all directions is conserved.
In such quantum billiards, the boundary conditions imposed on the  wave functions inside the billiards, associated with reflections at the boundaries, are accounted for semiclassically by the choice of extra phase factors ${\rm e}^{i\mu \pi}$ with $\mu=0,1$ refering to Neumann or Dirichlet boundary conditions, respectively. Once these extra phases are accounted for, the Berry-Tabor trace formula correctly describes the density of states by means of the torus foliation of the classical phase space. In this sense, one can say that the presence of Neumann/Dirichlet boundary conditions does not affect the integrability of the classical limit. 

The situation is different when the boundary conditions are of the mixed type. In a remarkable paper, Sieber et al.~\cite{Sieber1995} constructed the semiclassical trace formula for a billiard system in two dimensions with mixed boundary conditions. As we show now, proper identification of parameters allows one to identify the LL model for $N=2$ as formally identical to  such a 2D billiard system.

 Consider then the eigenstates $\psi_{n}(x,y)$ of a Hamiltonian describing a particle of mass $m$ moving inside a square domain of in the $x-y$ plane of linear size $L$, 
 \begin{align}
	\left(- \partial_{x}^2 -\partial_{y}^{2}\right)\psi_{n}(x,y)=E_{n}\psi_{n}(x,y)
\end{align}
 where we again choose units such that $\hbar=2m=1$. By imposing periodic boundary conditions in both directions,
 \begin{align}
     \psi_{n}(x,y)=\psi_{n}(x+L,y)=\psi_{n}(x,y+L)
 \end{align}
 the domain is then identified with a two-torus and the configuration variables $x,y$ as angles. 

 The system is obviously invariant under the exchange $x \leftrightarrow y$, thus separating the spectrum into sectors of even/odd  $\psi_{n}^{\pm}(x,y)=\pm\psi^{\pm}_{n}(y,x)$ behaviour under the exchange. As expected, in order to reach an analogy with the LL model, we will focus now on the even (bosonic) sector $(+)$. The dynamics now takes place in the fundamental domain, given simply by the triangular shape obtained by cutting the original square along $x=y$.

While in the absence of any other potential barrier the eigenstates corresponding to the even/odd sectors of the spectrum automatically satisfy Neumann/Dirichlet boundary conditions along the side $x=y$ of the resulting triangular billiard, the exchange symmetry is not affected by introducing a more general type of boundary condition. Following \cite{Sieber1995} we then impose mixed (Robin) type of condition
\begin{align}
\label{eq:Robin}
    \partial_{q}\psi^{+}_{n}\left. \left(X-\frac{q}{2},X+\frac{q}{2}\right)\right|_{q=0}=\kappa \psi_{n}^{+}(X,X)
\end{align}
where $X,q$ are longitudinal and transversal coordinates along the boundary. The apparent inconsistency between Eq.~(\ref{eq:Robin}) and the usual Neumann conditions for the bosonic sector (that would imply the vanishing of the left hand side) is solved by allowing for a {\it discontinuous} jump of the normal derivative across $x=y$. 

The full analogy with the $N=2$ LL model emerges then when we then rewrite the mixed boundary condition and the jump of the normal derivative by means of an extra delta potential 
\begin{align}
    V(x,y)=g\delta(x-y)
\end{align}
in the $q=y-x$ direction. A standard calculation shows that the presence of such delta line along $x=y$ together with the bosonic symmetry is exactly equivalent to the boundary condition, Eq.~(\ref{eq:Robin}) where $\kappa=g/2$. 

This equivalence of the $N=2$ LL model with the even sector of a 2-dimensional square billiard with periodic boundary conditions and a mixed boundary condition along the symmetry line, that can be studied with the methods of \cite{Sieber1995}, allows then for a non-trivial consistency check of the results presented here.

\section{V. Conclusions and further perspectives} 

We have shown how to extend the trace formula for quantized classical systems with integrable dynamics, pioneered by Berry and Tabor in the seminal papers \cite{Berry1976,Berry1977}, into the domain of \textit{quantum} integrable systems. Prime examples of the latter are many-body systems of interacting particles in one dimension that are solvable by means of Bethe ansatz techniques and lack a notion of classical limit in the unsual sense due to the short-range (contact) interactions. The resulting trace formula is obtained by using the structural similarity between the Bethe equations and EBK-type quantization conditions. This implies that such semiclassical expressions for the many-body density of states then could play a similar key role as its Berry-Tabor counterpart for elucidating important questions such as the emergence of Poisson-type spectral fluctuations in these models. We explicitly construct the many-body Berry-Tabor trace formula for the paradigmatic example of the Lieb-Liniger model. We show that it follows an expected progressively better agreement with exact numerically evaluated few-body energy levels upon increasing the number of formal periodic few-particle orbits.  The trace formula turns out to be powerful enough to exhibit the phenomenon of resurgence, now in the few-body spectrum.

Following the ideas presented here, important and experimentally relevant models, like the fermionic Hubbard model that also display quantum integrability, are now accessible to the powerful methods of semiclassical quantization. 
 

\section*{Acknowledgements}
\label{sec:acknowledgments}

We want to thank Quirin Hummel for his help in understanding the combinatorics leading to Eq.~(\ref{Hummel}), and an anonymous referee for interesting suggestions. We also acknowledge financial support from the Deutsche Forschungsgemeinschaft (German Research Foundation) through Ri681/15-1 within the Reinhart-Koselleck Programme.


\bibliographystyle{apsrev4-2}
\bibliography{library.bib}

\end{document}